\documentclass[reprint,
superscriptaddress,showpacs,
amsmath,amssymb,
aps,
pra,
floatfix
]{revtex4-1}

\usepackage{graphicx}
\usepackage{dcolumn}
\usepackage{bm}
\begin{document}

\preprint{APS/123-QED}

\title{Pulse propagation, population transfer and light storage in
  five-level media}

\author{G. Grigoryan}
\affiliation{Institute for Physical Research, Armenian National
  Academy of Sciences, Ashtarak-2, 0203, Armenia}
\affiliation{Russian-Armenian (Slavonic) University, Yerevan-51, 0051,
  Armenia}

\author{V. Chaltykyan}
\email{Deceased}
 
\affiliation{Institute for Physical Research,
  Armenian National Academy of Sciences, Ashtarak-2, 0203, Armenia}
\affiliation{Russian-Armenian (Slavonic) University, Yerevan-51, 0051,
  Armenia}

\author{E. Gazazyan}
\affiliation{Institute for Physical Research, Armenian National
  Academy of Sciences, Ashtarak-2, 0203, Armenia}
\author{O. Tikhova}
\affiliation{Institute for Physical Research, Armenian National
  Academy of Sciences, Ashtarak-2, 0203, Armenia}

\author{V. Paturyan}
\affiliation{Department of Computer Science,
  Maynooth University, Maynooth, Co Kildare, Ireland}
\email{misha@cs.nuim.ie}
\affiliation{School of Theoretical Physics, Dublin Institute for
  Advanced Studies, 10 Burlington Road, Dublin 4, Ireland }

\date{January 12, 2015}

\begin{abstract}
We consider adiabatic interaction of five-level atomic systems and
their media with four short laser pulses under the condition of all
two-photon detunings being zero. We derive analytical expressions for
eigenvalues of the system's Hamiltonian and determine conditions of
adiabaticity for both the atom and the medium. We analyse, in detail,
the system's behaviour when the eigenvalue with non-vanishing energy
is realized. As distinct from the usual dark state of a five-level
system (corresponding to zero eigenvalue), which is a superposition of
three states, in our case the superposition of four states does
work. We demonstrate that this seemingly unfavourable case
nevertheless completely imitates a three-level system not only for a
single atom but also in the medium, since the propagation equations
are also split into those for three- and two-level media
separately. We show that, under certain conditions, all the coherent
effects observed in three-level media, such as population transfer,
light slowing, light storage, and so on, may efficiently be realized
in five-level media. This has an important advantage that the light
storage can be performed twice in the same medium, i.e., the second
pulse can be stored without retrieving the first one, and then the two
pulses can be retrieved in any desired sequence.
\end{abstract}

\maketitle

\section{Introduction}

Coherent interaction of light signals with quantum systems attracted
considerable interest for their importance in both fundamental science
and practical applications. A prominent example of coherent
interactions is electromagnetically induced transparency (EIT)
\cite{fleischhauer2005electromagnetically,
  harris2008electromagnetically,scully1997quantum} which can be used
to eliminate the resonant absorption of a laser beam incident upon a
coherently driven medium with appropriate energy levels. EIT technique
allows controlled manipulations of the optical properties of atomic or
atom-like media via coupling them with signal and control fields. In
particular, it is possible to greatly slow down the optical (signal)
pulse \cite{hau1999light, budker1999nonlinear,kash1999ultraslow} and
even stop it to attain reversible storage and retrieval of information
\cite{fleischhauer2000dark,liu2001observation,phillips2001storage}.

Despite a huge number of publications, light storage remains in the
focus of attention of researchers, since it is one of key components
in optical (quantum) information processing \cite{lukin2003colloquium,
  fleischhauer2005electromagnetically, eisaman2005electromagnetically,
  chaneliere2005storage, kimble2008quantum,
  hammerer2010quantum}. Another application of coherent interactions
is controllable population transfer between the atomic levels and
constructing desired coherent superpositions of different states
\cite{bergmann1998coherent, kral2007colloquium,
  yatsenko2014detrimental}. These effects are also employed widely in
such fields of research as laser cooling of atoms, lasing without
inversion, new precision techniques of magnetometry, coherent control
of chemical reactions, and so on.

All the above-listed phenomena are comprehensively studied, both
theoretically and experimentally, for various three level systems and
their media \cite{novikova2012electromagnetically, li2013entanglement,
  heinze2013stopped, wesenberg2007scalable, yang2009all,
  saffman2010quantum,sargsyan2006dark}. Although multilevel atomic and
atom-like systems do not provide new physical principles in addition
to quantum interference and principle of superposition, they widen
essentially the possibilities of experimental realizations and
practical applications. The idea of a double-EIT (DEIT) regime is
introduced in \cite{lukin2000nonlinear} and modified in
\cite{petrosyan2002symmetric}. The laser cooling scheme for trapped
atoms or ions which is based on DEIT is discussed in
\cite{evers2004double}. DEIT in a medium, consisting of four level
atoms in the inverted-Y configuration is discussed in
\cite{harden2011demonstration}. DEIT in a ring cavity is studied in
\cite{huang2014double}. Enhanced cross-phase modulation based on DEIT
is reported in \cite{li2008enhanced}. Work \cite{zimmer2008dark}
examines dark state polariton formation in a four-level
system. Quantum memory for light via stimulated off-resonant Raman
process is considered in \cite{sheremet2010quantum} beyond the
three-level approximation. Work \cite{choi2014influence} proposes,
through numerical calculations, to use multilevel systems involving
hyperfine structure in problems of localization of excitations via
dark state formation in the EIT processes. Work
\cite{ottaviani2006quantum} investigates five-level atoms and media
driven by four light pulses in nonadiabatic regime. Two of four pulses
are assumed weak and treated as perturbation in the first order.  Work
\cite{chen2009detuned} observed experimentally off-resonance EIT-based
group delay in multilevel $D_{2}$ transition in rubidium. Enhancement
of EIT in a double-lambda system in cesium atomic vapor by specific
choice of atomic velocity distribution is observed in
\cite{scherman2012enhancing}. A scheme based on two sequential STIRAP
processes with four laser fields is proposed in
\cite{moller2007efficient} for measurement of a qubit of two magnetic
sublevels of the ground state of alkaline-earth metal ions. Another
topic where multilevel systems were used was generalization of the
notion of dark-state polariton \cite{joshi2005generalized} and
discussion of possibility to apply multilevel EIT to quantum
information processing.
\begin{figure}[ht]
  \centering
  \includegraphics[width=8.5cm]{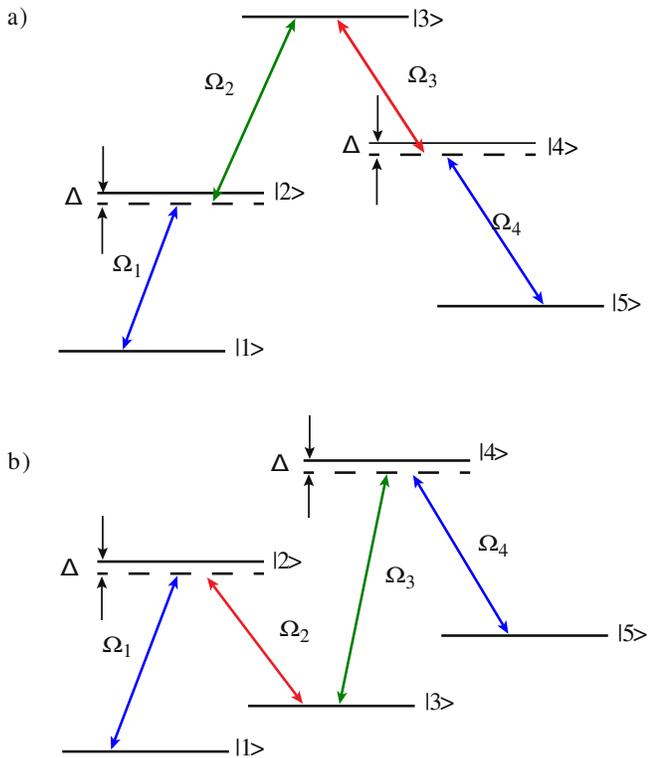}\\
  \caption{Five-level coupling schemes: (a) Extended
    $\Lambda$-scheme, (b) M-type scheme.}\label{1u1}
\end{figure}

In the present paper we study both analytically and numerically a
five-level atomic system interacting adiabatically with four
co-propagating laser pulses of different durations and different
sequences of turning on and off. We require that each laser pulse
interacts (be resonant) with only one of the adjacent transitions, and
assume all the two-photon detunings to be zero. Two examples of such
level diagrams are shown in Fig.~\ref{1u1}. Another example is the
ladder-system which may turn out to be rather useful for problems of
excitation of Rydberg states.

As distinct from all above-sited works, we concentrate on those
eigenstates of interaction Hamiltonian (see below) whose eigenvalues
are different from zero. We will show that these eigenstates are
similar to the well-known dark and bright states in a three-level
system. As distinct from the state in the M-system considered in
\cite{bergmann1998coherent}, the levels 2 and 4 are at interaction
with laser fields populated, but the population of level 3 remains
zero. We will demonstrate for our case that an efficient and more
flexibly controllable population transfer and light storage becomes
possible. We also study the advantages of this
technique. Specifically, we show the possibility of successive storage
of two pulses with their subsequent retrieval. The first pulse is
stored into the coherence $\rho_{51}$, which exactly reproduces, after
turning off the interaction, the shape of the pulse $\Omega_2$. Since
the coherence $\rho_{31}$ remains zero during all time of interaction,
the same medium can be used again for storage of another pulse.

The paper is organized as follows: In section \ref{sec2} we derive
eigenfunctions and eigenvalues of systems under consideration and
discuss the relevant cases. In section \ref{sec3} we study adiabatic
population transfer in five-level systems. Section \ref{sec4} derives
the equations of propagation and presents their analytical
solution. In the same section the regime of adiabaton is
demonstrated. In section \ref{sec5} we show the possibility to store
optical information in considered media. We conclude with a final
discussion in section \ref{sec6}.

\section{Eigenfunctions and eigenvalues of interaction Hamiltonian}
\label{sec2}
Consider a five-level atomic system as shown in Fig.~\ref{1u1}. Four
(in general) laser pulses are close to resonance with respective
transitions (Fig.~\ref{1u1}). Hamiltonian of interaction in
rotating-wave approximation, and under the assumptions that the
carrier frequencies of laser pulses are tuned near resonance with one
of the adjacent atomic transitions, and pulse durations are much
shorter compared to relaxation times in the system, has the following
form:
\begin{equation}\label{hamilt}
  H=\sum_i\sigma_{i,i}\delta_{i-1}-\biggr(\sum_i\sigma_{i,i+1}\Omega_i
  + h.c.\biggl),
\end{equation}
with the projection matrices $\sigma_{ij}=| i\rangle\langle j|$, the
Rabi frequencies $\Omega_i$ at transitions $i\rightarrow i$+1, and
$\delta_{i-1}$ representing ($i$-1)-photon detunings (with
$\delta_0=0$). The Rabi frequencies are assumed to be real and
positive. Phases, which can  vary during propagation, are included in
the single-photon detunings
($\Delta_i=\omega_{i+1,i}-\omega_i+\dot{\varphi}_i$, if
$\omega_{i+1,i}>0$ and
$\Delta_i=\omega_{i,i+1}-\omega_i+\dot{\varphi}_l$ if
$\omega_{i+1,i}<0$). Definition of multi-photon detunings depends on
the specific scheme of interaction. For an M-system (see
Fig.~\ref{1u1}(b)) the multi-photon detunings are
$\delta_2=\Delta_1-\Delta_2,\ \delta_3=\Delta_3+\delta_2,
\delta_4=\Delta_4-\delta_3$. For an extended $\Lambda$-system (see
Fig.~\ref{1u1}(a)), the multi-photon detunings are
$\ \delta_2=\Delta_1+\Delta_2, \delta_3=-\Delta_3+\delta_2,
\delta_4=-\Delta_4+\delta_3$. 

Eigenvalues of the Hamiltonian \eqref{hamilt} can easily be derived
analytically if all of the two-photon detunings are zero, i.e.  
\begin{equation}\label{2photon}
   \delta_2=0,  \delta_3-\delta_1=0,  \delta_4-\delta_2=0.
\end{equation}
For an M-system, these conditions mean equal single-photon detunings,
while for the extended $\Lambda$-system the single-photon detunigs
have equal absolute values, but differ in sign (see Fig.~\ref{1u1}).
When conditions \eqref{2photon} are met, one of five eigenvalues of
the Hamiltonian is $\lambda =0.$ The detailed calculations of
the remaining four eigenvalues are presented in Appendix A.

Consider now a special case, when the pulses $\Omega_1$ and $\Omega_4$
coincide by their temporal profiles (but the frequencies and phases of
pulses may be different).  In this case, the eigenvalues of the
Hamiltonian \eqref{hamilt} are
\begin{equation}\label{eq2}
\begin{gathered}
\lambda_0  =0,\\
\lambda_{1,3}  =\frac{1}{2}\biggr(\Delta\mp\sqrt{\Delta^2+
     4\Omega^2_1}\biggl),\\
\lambda_{2,4}  =\frac{1}{2}\biggr(\Delta\mp\sqrt{\Delta^2+
     4(\Omega^2_1+\Omega^2_2+\Omega^2_3)}\biggl)
\end{gathered}
\end{equation}
We note, that when the fields are turned off, we get
$\lambda_{1,2}\rightarrow 0$ and $\lambda_{3,4}\rightarrow\Delta.$ It
should be emphasized that the eigenvalues $\lambda_{1,3}$ depend upon
only the field $\Omega_1$ and coincide with the eigenvalues of a
two-level system, driven by field $\Omega_1$.  Similarly, the
eigenvalues $\lambda_{2,4}$ are equal to the eigenvalues of a
two-level system, driven by an effective field
$(\Omega^2_1+\Omega^2_2+\Omega^2_3)^{1/2}$.  Adiabatic evolution
requires the following conditions to be met (see Appendix A for
details):

\begin{equation}\label{eq3}
\begin{split}
 \Delta T & \gg1,\\
 \frac{(\Omega^2_2+\Omega^2_3)T}{\Delta} & \gg1,\\
 \frac{\Omega^2_1T}{\Delta} & \gg1
\end{split}
\end{equation}
with the duration $T$ of the shortest pulse. The first condition
mirrors the adiabaticity condition for a two-level system. The second
condition corresponds to the adiabaticity condition for a three-level
system. The third condition is only relevant in the time interval
where all pulses overlap (i.e., when $\Omega^2_2+\Omega^2_3\neq0$).

To write the eigenvectors corresponding to the eigenvalues $\lambda_1$
and $\lambda_2$ we introduce the following notations:
\begin{equation}\label{eq4}
\begin{gathered}
\Omega^2=\Omega^2_2+\Omega^2_3,\tan\theta=\frac{\Omega_2}{\Omega_3},\\
\tan{\Phi_1}=-\frac{\lambda_1}{\Omega_1},
\tan{\Phi_2}=-\frac{\lambda_2}{\Omega_1},\\
\tan{\Phi}=-\frac{\Omega}{\Omega_1}\cos{\Phi_2}
\end{gathered}
\end{equation}
Then, eigenvector corresponding to the eigenvalue $\lambda_1$ is
\begin{equation}\label{L1}
|\lambda_1\rangle=|\psi_1\rangle\cos{\theta}-|\psi_2\rangle\sin{\theta}
\end{equation}
where $|\psi_1\rangle$ and $|\psi_2\rangle$ are superposition states
of two-level systems $1\rightarrow2$ and $5\rightarrow4$:
\begin{equation}\label{eq6}
\begin{split}
|\psi_1\rangle=\cos{\phi_1}|1\rangle-\sin{\phi_1}|2\rangle\\
|\psi_2\rangle=\cos{\phi_1}|5\rangle-\sin{\phi_1}|4\rangle
\end{split}
\end{equation}
It is apparent that the eigenvector corresponding to the $\lambda_1$ does
not involve state $|3\rangle$ and is equal to the dark state of a
three-level $\Lambda$-system, if we replace the lower states by the
superposition states $|\psi_1\rangle$ and $|\psi_2\rangle$.

\begin{figure}[ht]
  \includegraphics[width=8.5cm]{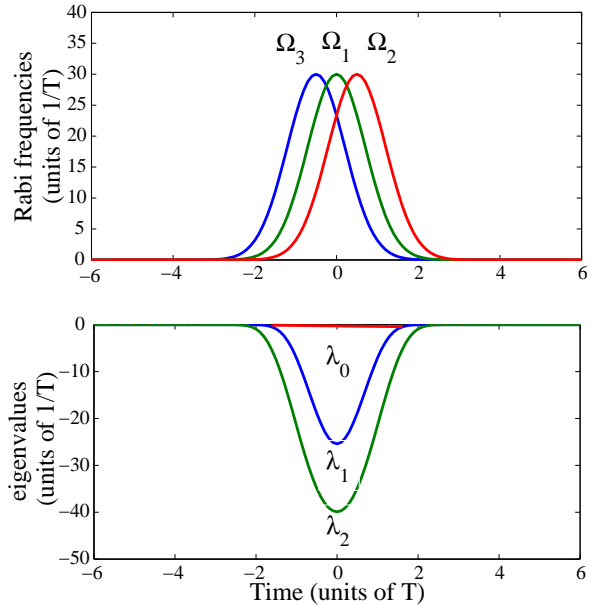}\\
  \caption{Time dependencies of the pulses (top) and
    the adiabatic energies (bottom) for considered systems. Shapes of
    all pulses are Gaussian. The single-photon detuning is
    $\Delta=10/T$. It is seen clearly that in the range of overlapping
    of the pulses, the adiabaticity of interaction is
    ensured.}\label{p4}
\end{figure}

\begin{figure}[ht]
   \begin{center}
  \includegraphics[width=8.5 cm]{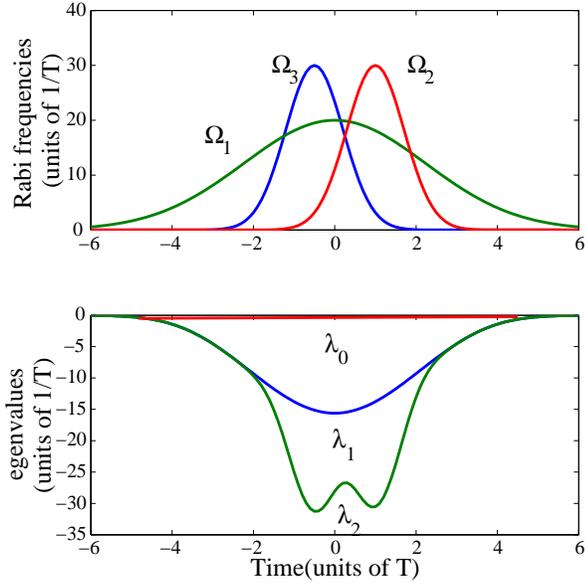}\\
  \end{center}
  \caption{Same as in Fig.\ref{p4}, but for another
    sequence of pulses.}\label{p5}
\end{figure}

Similarly, the eigenvector corresponding to the eigenvalue $\lambda_2$
yields
\begin{equation}\label{L2}
  |\lambda_2\rangle=|\psi'_1\rangle \cos{\Phi}\sin{\theta}-
       \sin{\phi}|3\rangle +|\psi'_2\rangle \cos{\Phi}\cos{\theta},
\end{equation}
where
\begin{equation}\label{eq8}
\begin{split}
|\psi'_1\rangle=\cos{\Phi_2}|1\rangle-\sin{\Phi_2}|2\rangle\\
|\psi'_2\rangle=\cos{\Phi_2}|5\rangle-\sin{\Phi_2}|4\rangle
\end{split}
\end{equation}
As in the previous case, the eigenvector $|\lambda_2\rangle$ is equal
to that of the bright state of a three-level $\Lambda$-system
\cite{klein2007robust,grigoryan2009theory}, if we replace the lower
states by superposition states $|\psi'_1\rangle$ and
$|\psi'_2\rangle$.  The time behavior of eigenvalues $\lambda_i$ in
the special case above for different pulse sequences is demonstrated
in Fig.~\ref{p4} and Fig.~\ref{p5}

\section{Population transfer}
\label{sec3}
As follows from the expressions \eqref{L1} and \eqref{L2} the five-level
system imitates the three-level lambda system.
\begin{figure}[ht]
    \begin{center}
   \includegraphics[width=8.5cm]{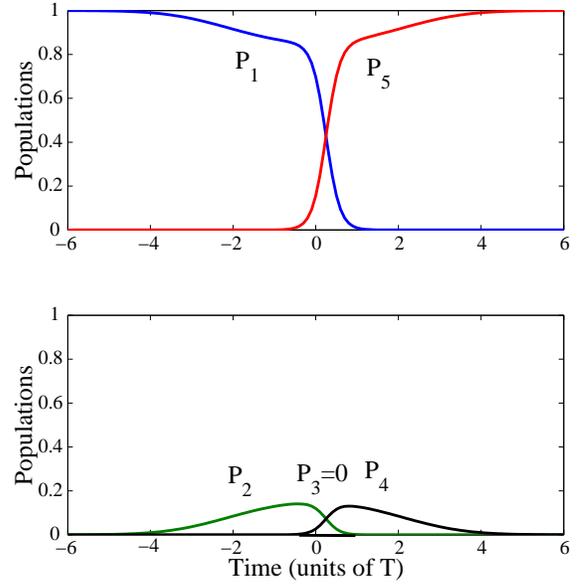}\\
  \end{center}
  \caption{Dynamics of the population transfer from
    initial state $|1>$ to final state $|5>$ if the atom is in state
    $|\lambda_1>$ and the pulse sequence is the same as in
    Fig.~\ref{p5}. ($P_i=|<i|\lambda_1>|^2$). The single photon
    detuning is $\Delta=20/T $.}\label{p8}
\end{figure}
Thus, we can use the state $|\lambda_1\rangle$ to transfer the system
from state $|1\rangle$ to state $|5\rangle$ by a STIRAP-like process,
driven by the pulse sequence introduced above (see Fig.~\ref{p8}). In
contrast to a simple three-level $\Lambda$-system, during the
interaction some transient population shows up in the intermediate
levels $|2\rangle$ and $|4\rangle$ of the five-level system. However,
these transient populations are very small, if the one-photon detuning
is sufficiently large, but still satisfies the adiabaticity condition
(3). It should be noted that the condition of large one-photon
detuning is not very crucial for the population transfer. The
dynamics of populations in the described case is shown in
Fig.~\ref{p8}.

\begin{figure}[ht]
    \begin{center}
  \includegraphics[width=8.5cm]{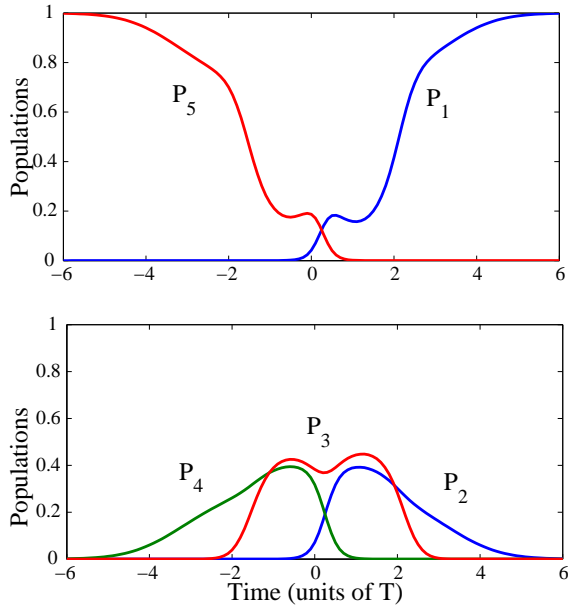}\\
  \end{center}
  \caption{Dynamics of the population transfer from
    initial state $|5>$ to final state $|1>$ if the atom is in state
    $|\lambda_2>$. The pulse sequence and parameters are the same as
    in Fig.~\ref{p8} ($P_i=|<i|\lambda_2>|^2$).}\label{p6}
\end{figure}

Similarly, the state $|\lambda_2\rangle$ is analogous to the bright
state of a lambda-system and we can use these states for adiabatic
transfer from state $|5\rangle$ to state $|1\rangle$ by a
b-STIRAP-like process \cite{klein2007robust,grigoryan2009theory}
driven by the pulse sequence in Fig.~\ref{p5}, because the state
$|\lambda_2\rangle$ is not realized with the pulse sequence of
Fig.~\ref{p4}, according to the definition of the angle $\Phi$. We
emphasize that the STIRAP-technique is applicable for both schemes of
pulse sequence in Figs.~\ref{p4} and \ref{p5}.  The dynamics of
populations in the b-STIRAP case are demonstrated in Fig.~\ref{p6}.
Note that the two eigenstates $|\lambda_1\rangle$ and
$|\lambda_2\rangle$ render the five-level system, driven by a
considered pulse sequence, fully reversible. Thus, we can transfer
atomic population from state $|1\rangle$ to state $|5\rangle$ by a
STIRAP-like process and from state $|5\rangle$ to state $|1\rangle$ by
a b-STIRAP-like process with the same sequence of pulses.

\section{Medium of atoms}
\label{sec4}
Now we move from a single-atom case to that of a medium consisting of
the described atoms. We start from the well-known truncated Maxwell
equation in running coordinates $x$, $\tau=t-x/c$.:
\begin{equation}\label{Maxsvel}
    \frac{\partial E_i}{\partial x} = i \frac{2 \pi  \omega_i }{c} Nd_i
\end{equation}
Here $E_i$ are the complex amplitudes of electric fields of the
pulses, N is the number density of medium atoms, and $d_i$ are the
amplitudes of induced dipole moments of each individual atom at a
frequency $\omega_i$, $\langle\psi| d|\psi\rangle=\sum d_i
exp(-i\omega_it)+c.c.$. These amplitudes can be expressed in terms of
the amplitudes of atomic populations $b_{i}$ of bare states and the
matrix elements of the dipole moment $\langle i|d|i+1\rangle$:
$d_i=b^{*}_{i}b_{i+1}\langle i|d|i+1\rangle$ if $\omega_{i+1,i}>0$ and
$d_i=b_{i}b^{*}_{i+1}\langle i|d|i+1\rangle$ if
$\omega_{i+1,i}<0$. The coefficients $b_{i}$ are determined by the
non-stationary Schroedinger equation with Hamiltonian \eqref{hamilt}.

Separating real and imaginary parts in the truncated equation of
propagation, differentiating the equation for the phase with respect
to time, and combining the obtained equations with the Schroedinger
equation, we obtain in the general case a self-consistent system of
equations describing variation of frequencies (one-photon detunings)
and intensities (Rabi frequencies) of pulses during propagation in
medium. For example, in the case of medium consisting of M-type atoms
we obtain:
\begin{equation}\label{M-system}
\begin{split}
 \frac{\partial\Omega^2_1}{\partial x} &=
   q_1\frac{\partial|b_1|^2}{\partial\tau}\\
 \frac{\partial\Omega^2_2}{\partial x} &=
   -q_2\frac{\partial(|b_1|^2+|b_2|^2)}{\partial\tau}\\
 \frac {\partial\Omega^2_3}{\partial x} &=
   -q_3\frac{\partial(|b_4|^2+|b_5|^2|)}{\partial\tau}\\
 \frac{\partial\Omega^2_4}{\partial x} &=
   q_5\frac{\partial|b_5|^2}{\partial\tau}\\
 \frac{\partial\Delta_i}{\partial x} &=
   q_i\frac{\partial}{\partial\tau}\frac{Re(b^*_i b_{i+1})}{\Omega_i}
\end{split}
\end{equation}
where $q_i=2 \pi N \omega_i |d_{i,i+1}|^2/c\hbar$ is the propagation
constant.  In the case of an extended $\Lambda$-system
(Fig.~\ref{1u1}a) the equations remain essentially the same, but we
must change the signs of the rhs in the second and third equations.
As follows from equations \eqref{M-system}, during propagation in the
medium not only the shapes of pulses may vary essentially, but also
the conditions for detuning of resonances may be violated. Variations
of resonance detunings are caused by processes of self-phase
modulation (parametric broadening of pulse spectrum
\cite{shen1984principles}). The modification of the shapes of pulses
is caused both by the nonlinear group velocity (which can result in
formation of shock wavefronts \cite{grischkowsky1973observation}) and
by energy transfer between the pulses (which can lead to full
depletion of one of pulses \cite{chaltykyan2003dark}). It is, however,
obvious that all these processes are proportional to the length of
propagation. Hence, if the optical length of the medium is
sufficiently short, the variations of detunings and intensities can be
negligibly small.

Since it is only the time derivatives that enter the right-hand
sides of the equations \eqref{M-system}, we can use expressions \eqref{L1}
and \eqref{L2} for the atomic amplitudes in these equations and this
will be equivalent to allowance for the first non-adiabatic
corrections.  Correspondingly, the conditions of smallness of rhs of
\eqref{M-system} serve as a criteria of insignificance of changes in
spatial and temporal characteristics of pulses and thus a criteria of
adiabaticity of the interaction in the medium.  For simplicity we will
restrict ourselves to the case of equal oscillator strengths in all
transitions (in the case of different oscillator strengths, we can
proceed as in the three-level system \cite{grigoryan2009short}).  In
case of the state $|\lambda_1\rangle$ we now obtain from
\eqref{M-system}

\begin{equation}\label{osn}
\begin{gathered}
  \frac{\partial(\Omega^2_1+\Omega^2_4)}{\partial x}=
  q\frac{\partial\cos^2 \Phi_1}{\partial \tau },\\
  \frac{\partial(\Delta_1+\Delta_4)}{\partial x}=
  -q\frac{\partial}{\partial\tau}\frac{\cos(2\Phi_1)}{\Delta},\\
  \frac{\partial\Delta_2}{\partial x}=
  -\frac{\partial\Delta_3}{\partial
  x}=0,\\
\frac{\partial\Omega^2}{\partial x}=0,\\
\frac{\partial\theta}{\partial x}+
\frac{q}{\Omega^2}\frac{\partial\theta}{\partial \tau}=0
\end{gathered}
\end{equation}

\begin{figure}[ht]
  \includegraphics[width=8.5cm]{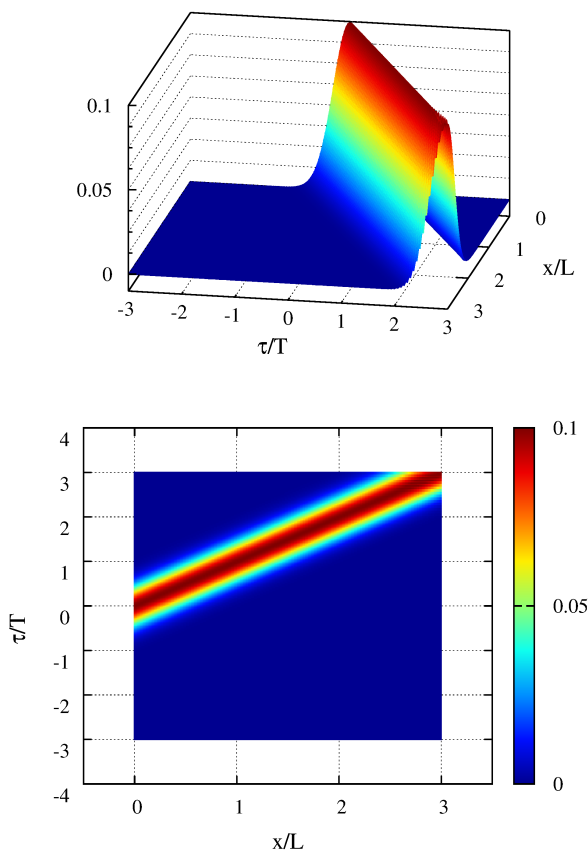}\\
   \caption{Distortion-free propagation of the signal
     pulse $\Omega_2$ at the subluminal group velocity. Shapes of
     pulses are chosen to be
     $\Omega_1T=\Omega_4T=\Omega_3T=30e^{(-0.2(\tau/T)^2)}$,
     $\Omega_2T=0.1e^{(-5(\tau/T)^2)}$ (linear case). The
     single-photon detuning is $\Delta=100/T$. The scaled length
     $L=\Omega^2_0 (\tau=0)T/q$ and the time delay in medium is
     $\Delta t/T=x/L$}\label{adiabaton}
\end{figure}

We emphasize that the system of equations \eqref{osn} has an
interesting and important peculiarity. Propagation of fields
$\Omega_2$ and $\Omega_3$ occurs independent of $\Omega_1$ and
$\Omega_4$ and is described by propagation equations for
three-level-atom medium in conditions of dark-state formation
\cite{fleischhauer2005electromagnetically}. The fields $\Omega_1$ and
$\Omega_4$ are described by propagation equations for two-level-atom
medium \cite{grischkowsky1973observation}. This peculiarity is
important because both problems are studied in sufficient detail in
the literature and have analytical solutions. In particular, we can
realize all phenomena taking place in the usual lambda systems with
the three-level 2-3-4 system which is supported by two-level 1-2 and
4-5 systems pumping level 2 and depleting level 4, respectively.  As
an example, we obtain, in the considered five-level system,
propagation of the adiabaton \cite{grobe1994formation} in the
five-level system. Fig.~\ref{adiabaton} visualizes this phenomenon
(details in figure caption).


Equations \eqref{osn} are valid if state $|\lambda_1\rangle$ is formed
on entire length of the medium. This requires fulfillment of two
conditions: i) the detunings $|\Delta_i|=\Delta$ for all $i$ and Rabi
frequencies $\Omega_1=\Omega_4$ and ii) the adiabaticity of
interaction in all of the medium. Let us examine when these conditions
are met. Equations \eqref{osn} show that detunings $\Delta_2$ and
$\Delta_3$ are preserved during propagation (as they should be in a
three-level system), whereas $\Delta_1$ and $\Delta_4$ can vary with
propagation length because of self-phase modulation (as in two-level
system), but, as shown in \cite{grischkowsky1973observation}, these
variations can be neglected if we limit the length by
\begin{eqnarray}\label{eq12}
\frac{q x}{\Delta}\frac{1}{\Delta T}\ll 1,
\end{eqnarray}
On the same length we can take $\Omega_1=\Omega_4$ (adiabatic
approximation for two-level system). As follows from the results of
cited works the adiabaticity of interaction in two-level system breaks
at the lengths when $(qx/\Delta^2T)\sim1$, whereas the interaction
adiabaticity in three-level medium does not break at all. Another
condition imposed on the length requires non-depletion of pump pulse
in three-level medium for an effective population
transfer\cite{chaltykyan2003dark}
\begin{eqnarray}\label{eq13}
\frac{q x}{\Delta}\frac{\Delta}{\Omega^2 T}\sim1
\end{eqnarray}
It follows from \eqref{eq12} and \eqref{eq13} that the influence of
medium is determined by the factor $qx/\Delta$, times the adiabaticity
conditions for a single atom. This means that it is sufficient to
require the medium parameter $qx/\Delta$ to not exceed unity by
much. If we express this parameter in terms of the linear coefficient
of absorption of medium $\alpha_0$, we obtain restriction for the
optical length in the form:
\begin{equation}\label{alpha}
\frac{qx}{\Delta}=\alpha_0 x \frac{\Gamma}{\Delta}\sim 1
\end{equation}
with $\Gamma$ being the maximum  of the relevant widths.

So, in the case of a large one-photon detuning the length of
adiabaticity of interaction can exceed the length of linear absorption
in medium several times.
On this length equations \eqref{osn} can be solved analytically and
the solution has the form:
\begin{equation}\label{solution}
\begin{gathered}
\Omega_1=\Omega_4=\Omega_{10}(\tau), \\
\Omega_2=\Omega_0(\tau)\sin\theta_0(\xi(x,t)),  \\
\Omega_3=\Omega_0(\tau)\cos\theta_0(\xi(x,t))
\end{gathered}
\end{equation}
where $\Omega_{10}$, $\Omega_0$, $\theta_0$ are the boundary
conditions given at the entrance of the medium and $\xi(x,t)$is an
implicit function defined by the following expression
\begin{equation}\label{sol2}
\int^{\tau}_\xi \Omega^2_0dt=qx
\end{equation}
We note, that all the above is true only if the dressed state
$|\lambda_1\rangle$ is realized in the medium. In case then, as a
result of an adiabatic interaction, the other dressed state (for
example $|\lambda_2\rangle$) is realized, the propagation equations
\eqref{M-system} are no longer split, and finding their solution
requires an additional investigation.
\section{Light storage}
\label{sec5}
It follows from the solution \eqref{solution} that, after turning off
all pulses, the coherence $\rho_{15}$ induced by these pulses remains
in the medium (like in three level system):
\begin{equation}
\rho_{15}=-\sin\theta(\xi)\cos\theta(\xi)
\end{equation}
where function $\xi(x)$ is defined by the following expression:
\begin{equation}\label{ksi}
\int^{\infty}_\xi \Omega^2_0dt=qx
\end{equation}

\begin{figure}[ht]
    \begin{center}
   \includegraphics[width=8.5cm]{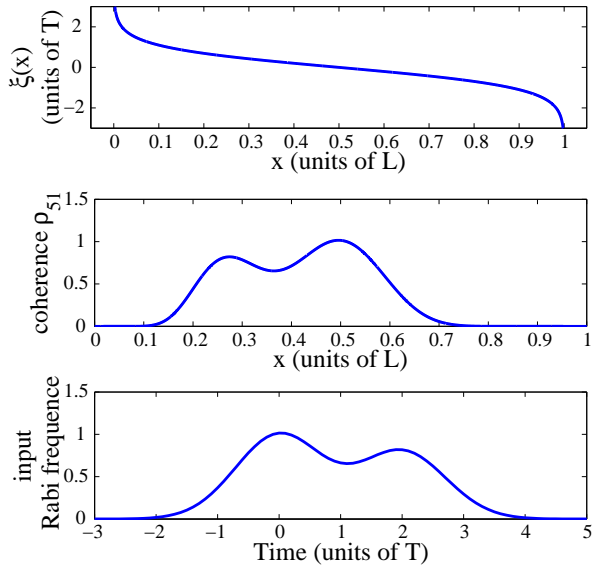}\\
  \end{center}
  \caption{The $x$-dependence of $\xi(x)$-function and
    the spatial distribution of coherence $\rho_{51}$ after the
    interaction is switched off (top) and the temporal profile of the
    pulse $\Omega_2$ at the medium input (bottom).}  \label{p7}
\end{figure}

\begin{figure}[ht]
   \includegraphics[width=8.5cm]{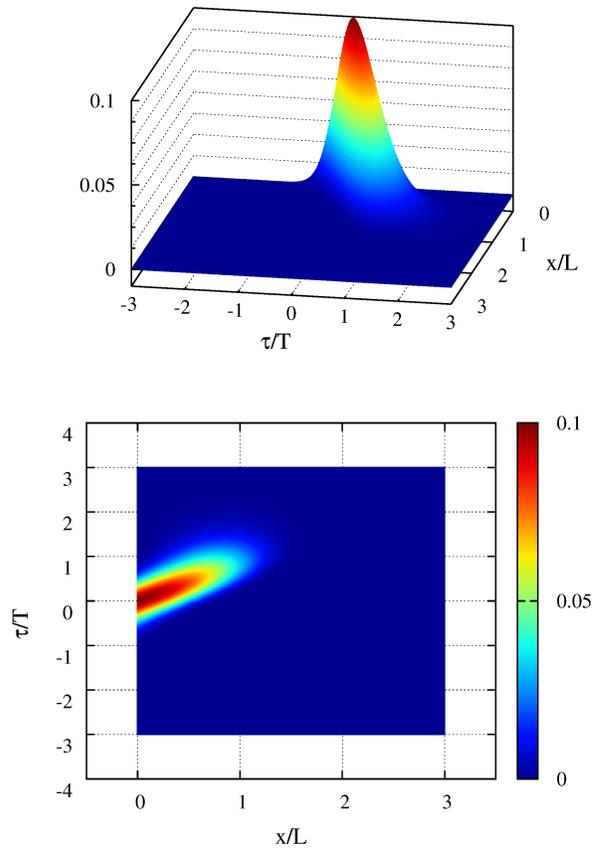}\\
   \caption{Propagation of the signal pulse $\Omega_2$
     (light storage). Shapes of pulses are chosen to be
     $\Omega_1T=\Omega_4T=30e^{(-3(\tau/T)^2)}$,
     $\Omega_3T=30e^{(-(\tau/T)^2)}$,
     $\Omega_2T=0.1e^{(-5(\tau/T)^2)}$. The single-photon detuning is
     $\Delta=100/T$. The group velocity
     $u=c/(1+qc/\Omega^2_3)\rightarrow0$ when
     $\Omega^2_3\rightarrow0$}\label{figure_light_storage}
\end{figure}

Fig.~\ref{p7} shows x-dependence of the $\xi$-function and coherence
$\rho_{15}$ after all pulses are turned off, together with the input
shape of the probe pulse. The figure demonstrates that the distribution of
coherence along $x$ mirrors the arbitrary $t$-shape of the probe pulse
at medium entrance.  It is also apparent that the $\xi$-function has
two asymptotes, $x=0$ and $x=x_{max}$. Existence of the maximal length
of medium, i.e., the length where probe pulse disappears, is the
essence of light storage phenomenon. For realization of storage (and
mapping of $t$-dependence onto $x$-distribution) the medium must be
not shorter than $x_{max}$.  It follows from \eqref{ksi} that the
maximal length is representable in the form $ x_{max}N=N_{ph}$, where
$N$ is the number density of resonant atoms in medium and $ N_{ph}$ is
the overall photon fluxes in control and probe pulses at the medium
input:
\begin{equation}\label{flux}
N_{ph}=\int^{\infty}_{-\infty}(cE_{p0}^2/\hbar\omega_p
  +cE_{s0}^2/\hbar\omega_s)dt
\end{equation}
Thus, in order to write completely a light pulse into a medium, it is
necessary that the number of atoms interacting with radiation be
comparable with the total number of relevant photons.We emphasize that
$x_{max}$ does not depend on $\Omega_1$ and $\Omega_4$ (the latter
enters only the adiabaticity condition). Note that in linear
approximation in $\Omega_2$ we can construct a dark-state polariton
similar to that in lambda system \cite{fleischhauer2000dark}.

Fig.~\ref{figure_light_storage} shows, by means of numerical solution
of corresponding equations with the use of Lax-Wendroff method
\cite{laxwendrof1960,trefethen}, the process of writing of a light
pulse into a medium.  The described process of light storage is more
visualisable in the generalized lambda-system, but has no principle
advantages as compared to the usual lambda system. In contrary, the
M-system is much more interesting because it enables double storage,
i.e., we can write two different pulses, one after another with
possibility of subsequent retrieval in any desired succession. Indeed,
during the whole interaction time (and also after the first writing),
the coherence $\rho_{31}$ remains zero and the population of the level
1 is close to unity (in linear approximation in $\Omega_2$). This
means that the same medium is ready for the usual lambda-storage of
the second pulse. For example, we can write the pulse $\Omega_1$ into
the same medium. For this purpose the pulses $\Omega_1$ and $\Omega_2$
should be divided into two beams before the first storage attempt. The
weak portion of $\Omega_1$ and the strong portion of $\Omega_2$ should
be sent to a delay line, to be used for the second storage (using
$\Omega_2$ as a control pulse).

\section{Conclusion}
\label{sec6}
We considered the behavior of a five-level atomic system and a medium
of such systems driven by four laser pulses of different amplitudes
and frequencies. We showed for such a system the possibility of
analytical determination of system eigenfunctions and eigenvalues in
case where all two-photon detunings are zero. We have obtained that,
in addition to the traditional zero eigenvalue, there exists a
non-zero one, for which the propagation equations in the medium are
split into the equations for two- and three-level system media, i.e.,
two of the four laser pulses travel independently of two other. This
splitting is caused by the fact that in this case the five-level
system reduces to a certain effective ``lambda''-system whose ground
states are superpositions of two states.  We derive the dressed states
and dressed energies of the system, as well as conditions for
adiabatic evolution, and show that the length of medium where
adiabaticity is preserved exceeds several times the linear absorption
length. We show that adiabatic passage permits reversible transfer of
atomic population from an initial to a target state, and back
again. The obtained mechanism of the population transfer may be
employed for excitation of Rydberg states in atoms.We analysed the
traveling of pulses in the medium and obtained, in particular,
adiabaton (distortion-free) propagation at the group velocity lower
than $c$. Also the process of information storage in five-level medium
was examined. We propose a possibility of double storage of light
pulses in the same medium with subsequent retrieval of the two stored
pulses in desired sequence. We note that the relaxation processes have
not been taken into account throughout the work. Allowance for these
processes requires separate investigation. Finally we note that the
considered five-level systems can experimentally be realized in a
number of media, such as hyperfine structures of D-lines of
alkali-metal atoms, in optical transitions of rare-earth-ion
impurities in crystal matrices, in rovibrational levels of different
electronic states in molecules, in problems of population transfer in
entangled three two-level atoms and so on.

\section*{Acknowledgments}
The research leading to these results has received funding from
Seventh Framework Programme FP7/2007-2013/ under REA Grant Agreement
No. 287252 and 295025. We acknowledge additional support from the
IRMAS International Associated Laboratory (CNRS-France SCS-Armenia).

\bigskip

\section*{Appendix A}
Equation for eigenvalues of the interaction Hamiltonian, i.e., the
equation $det(H -\lambda I)=0$, has, under conditions
$\delta_2=\delta_4=0$ and $\delta_1=\delta_3=\Delta$, the following
form:
\begin{equation}
  \lambda^2(\lambda-\Delta)[\lambda
    (\lambda-\Delta)+\Omega^2_s]+V^4\lambda =0
\end{equation}
where $\Omega^2_s=\Omega^2_1+\Omega^2_2+\Omega^2_3+\Omega^2_4$ and
$V^4=\Omega^2_2\Omega^2_4+\Omega^2_1\Omega^2_3+\Omega^2_1\Omega^2_4$. With
the notation $x=\lambda(\lambda-\Delta)$ the equation above becomes
$\lambda[x^2-\Omega^2_s x + V^4]=0$ and the eigenvalues are obtained
directly:

\begin{equation}
\begin{gathered}
\lambda_0=0,\\
\lambda_{3,1}=\frac{1}{2}[\Delta\pm(\Delta^2+4x_1)^{1/2}],\\
\lambda_{4,2}=\frac{1}{2}[\Delta\pm(\Delta^2+4x_2)^{1/2}]
\end{gathered}
\end{equation}
where $x_{2,1}=(1/2)[\Omega^2_1\pm(\Omega^4_s-4V^4)^{1/2}]$. We note
that the condition $\Omega^4_s\geq 4V^4$ is always met.

Conditions of interaction adiabaticity for a single atom,
$|\lambda_i-\lambda_j|T\gg1$ for any $i\neq j$ with $T$ being the time
of interaction, leads to following requirement imposed on the
parameters of pulses.

\begin{equation}
\begin{split}
\frac{(x_2-x_1)T}{(\Delta^2+4x^2_2)^{1/2}} & \gg1,\\
(\Delta^2+4x_1)^{1/2}T & \gg 1,\\
\frac{x_{1,2}T}{(\Delta^2+4x_{1,2})^{1/2}} & \gg 1
\end{split}
\end{equation}
Note that the last condition can be fulfilled only for $V^4\neq0$,
i.e., in the range of overlapping of pulses.

\bibliography{new}{}

\end{document}